\documentclass[a4paper,conference]{IEEEtran}
\usepackage[sorting=none]{biblatex}
\usepackage{bbm}
\usepackage{subcaption}

\usepackage{graphicx}
\graphicspath{ {./images/} }

\usepackage{amssymb}

\begin{document}

\title{Analysis of Augmentations for Contrastive ECG Representation Learning}

\author{\IEEEauthorblockN{Sahar Soltanieh$^1$, Ali Etemad$^{1,2}$, Javad Hashemi$^3$}
\IEEEauthorblockA{$^1$Dept. ECE, $^2$Ingenuity Labs Research Institute, $^3$School of Computing \\
Queen's University, Kingston, Canada\\
\{sahar.soltanieh, ali.etemad, javad.hashemi\}@queensu.ca}}

\maketitle

\begin{abstract}
This paper systematically investigates the effectiveness of various augmentations for contrastive self-supervised learning of electrocardiogram (ECG) signals and identifies the best parameters. The baseline of our proposed self-supervised framework consists of two main parts: the contrastive learning and the downstream task. In the first stage, we train an encoder using a number of augmentations to extract generalizable ECG signal representations. We then freeze the encoder and finetune a few linear layers with different amounts of labelled data for downstream arrhythmia detection. We then experiment with various augmentations techniques and explore a range of parameters. Our experiments are done on PTB-XL, a large and publicly available 12-lead ECG dataset. The results show that applying augmentations in a specific range of complexities works better for self-supervised contrastive learning. For instance, when adding Gaussian noise, a sigma in the range of 0.1 to 0.2 achieves better results, while poor training occurs when the added noise is too small or too large (outside of the specified range). A similar trend is observed with other augmentations, demonstrating the importance of selecting the optimum level of difficulty for the added augmentations, as augmentations that are too simple will not result in effective training, while augmentations that are too difficult will also prevent the model from effective learning of generalized representations. Our work can influence future research on self-supervised contrastive learning on bio-signals and aid in selecting optimum parameters for different augmentations. 
\end{abstract}


\section{Introduction}
An electrocardiogram (ECG) is a non-invasive and effective tool for measuring the electrical activity of the heart \cite{1}. ECG signals convey valuable information about the functionality and condition of the heart, including heart rate, blood pressure, and heart diseases. In recent years, deep learning solutions have been used to create effective and robust ECG-based diagnostics and therapeutic tools. Examples include the use of deep neural networks for detection of cardiovascular diseases such as different types of Arrhythmia \cite{andersen2019deep, isin2017cardiac, sannino2018deep, pyakillya2017deep, kadbi2006classification, sarkar2021detection}.

Supervised deep learning methods often require large amounts of labelled data to perform optimally. However, this is not always the case for many physiological data such as ECG recordings. The tedious nature of the labeling process and the required highly specialized domain knowledge makes the task extremely costly. In addition, some conditions such as long QT \cite{schwartz1975long} are relatively rare among the population, making the collection of such data even more difficult. Also, in many cases, the fluidity of medical definitions leads to high inter-observer variability and a lack of a gold standard (for instance the definition of type 3 long QT \cite{makita2008e1784k, hajimolahoseini2018ecg}). In such cases, classical \textit{supervised} deep learning methods may counter challenges and eventually lead to sub-optimal performance. 

\begin{figure}
    \begin{center}
    \includegraphics[width=1\columnwidth]{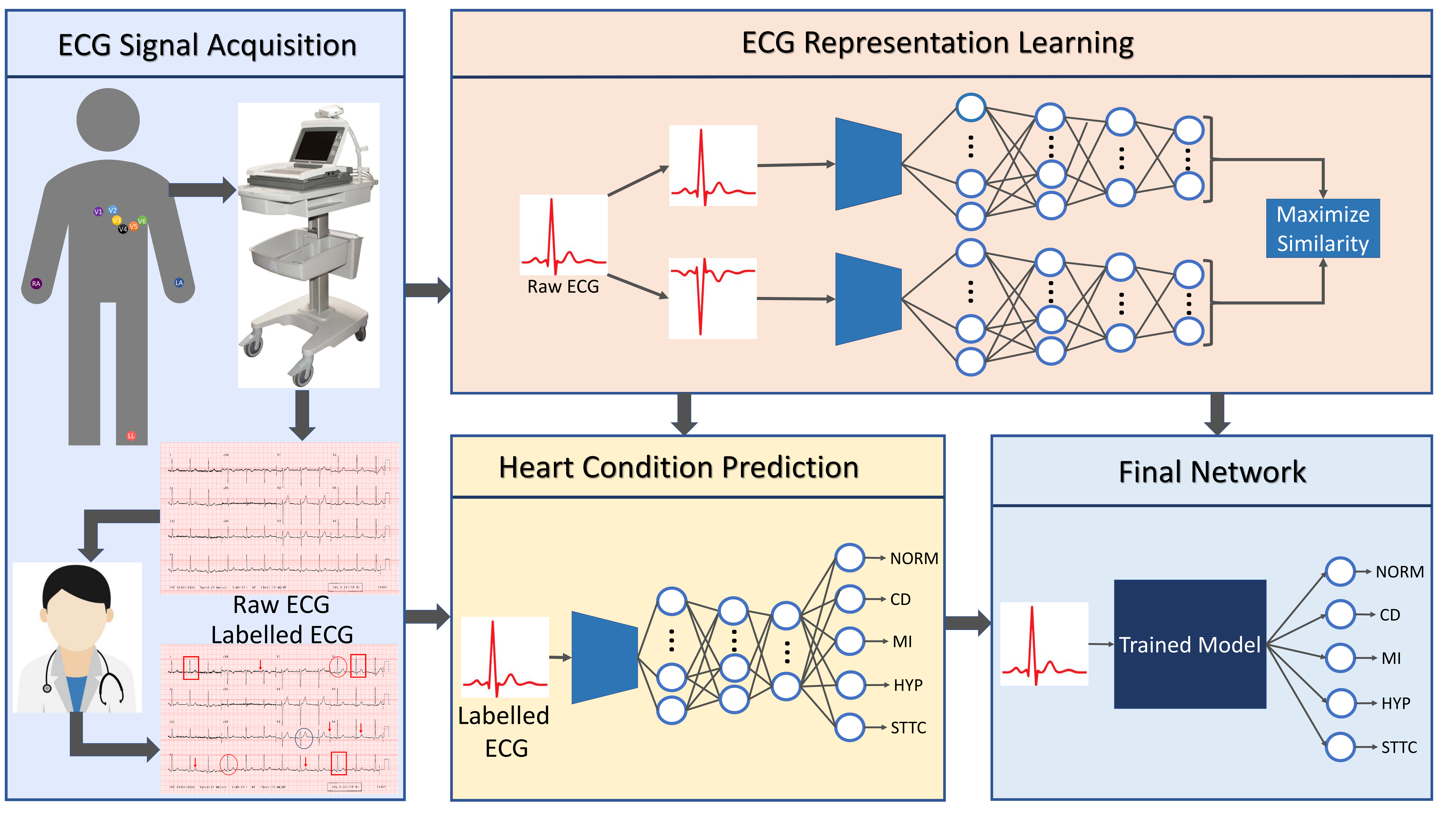}    
    \end{center}
\caption{An overview of a self-supervised ECG representation learning system for arrhythmia detection.}
\label{fig:pic}
\end{figure}

Self-supervised learning is a feature learning technique for overcoming the challenges related to the scarcity of `labelled' data \cite{joulin2016bag, noroozi2016unsupervised, liu2016recurrent, zhang2016colorful}. This relies on the generation of pseudo-labels from various augmentations for data during training instead of using pre-annotated data with ground truth labels. Thanks to their ability to generate a variety of different augmentations with varying degrees of difficulty, self-supervised learning often results in learning more generalized representations \cite{wang2017transitive}, making it highly accurate and robust toward different variations in the training data. 
Contrastive learning is a form of self-supervision in which a data sample is \textit{augmented} to create positive pairs, while other samples are considered negative pairs \cite{oord2018representation, chen2020simple, khosla2020supervised}. These positive and negative pairs are then used to train a Siamese style network and learn representations where the positive pairs maintain close proximity while the negative pairs are situated far apart in the feature space. Variations of contrastive methods in which negative pairs are no longer required for training have also been recently proposed \cite{chen2021exploring}.
Prior works have shown the the choice of `\textit{type}' and `\textit{intensity}' of augmentations used for contrastive self-supervised learning can have a considerable impact on the model performance \cite{tian2020makes, wang2021contrastive}. Thus, this paradigm poses a new set of challenges in that the optimum parameters regarding the augmentations need to be carefully designed and incorporated into the self-supervised training pipeline.

In this paper we design and implement a contrastive self-supervised framework for ECG-based arrhythmia detection (see Figure \ref{fig:pic}). First we contrastively train a deep convolutional neural network (CNN) encoder to learn representations from input ECG signals obtained from the PTB-XL dataset \cite{wagner2020ptb}. We then freeze the encoder and finetune a few linear layers with different amounts of labelled samples for downstream arrhythmia classification. Our goal is to provide a comprehensive analysis of the impact of different augmentations and their parameters towards contrastive ECG learning. To this end, we experiment with 7 different commonly used augmentations and systematically study their impact across different intensity parameters. We identify that the addition of Gaussian noise, permutation, and time warping yield strong results when specific ranges of parameters are used. We also experiment with different encoder architectures to ensure that our findings generalize well across different CNN encoders. 

\section{Related Work}

\subsection{Supervised ECG Learning}
There has been extensive research on ECG classification for different tasks. Researchers have used intrinsic features of ECG in classical machine learning techniques for ECG classification \cite{martis2009two, yu2008integration, tang2014classification}. In recent years, with the evolution of deep learning models and the creation of relatively large ECG datasets, ECG classification using deep learning models has become prominent \cite{zhang2017patient, ubeyli2010recurrent, wang2019global}. In this section, we provide a summary of some key works in the field of ECG classification, ranging from past classical machine learning-based algorithms to more recent deep learning-based methods.

In classical machine learning approaches, researchers largely relied on important ECG features. These features include the PQRST complex and their time intervals, such as the RR interval \cite{dallali2011fuzzy, ozbay2006fuzzy}, QRS duration \cite{tang2014classification, khandoker2008support}, QT interval \cite{martis2009two, karimipour2014real, yu2008integration}, as well as statistical and morphological features \cite{dilmac2015ecg, ince2009generic, de2004automatic, ye2012heartbeat, christov2006comparative}. A comprehensive survey in this domain can be found in \cite{jambukia2015classification}.

When it comes to deep learning solutions, we first present the prior work on ECG representation learning using fully supervised approaches based on two main categories of neural networks, namely convolutional neural networks (CNNs) and recurrent neural networks (RNNs). In \cite{kachuee2018ecg}, an ECG heartbeat classifier using a 1D CNN model was proposed. This classifier achieved strong performance for a five class arrhythmia classification. In \cite{li2018patient} a Generic Convolutional Neural Network was first trained on a large number of ECG signals. Then, to learn distinctive features of each patient's ECG, the network was finetuned on the patient's ECG signals individually. In \cite{baloglu2019classification}, an accurate and sensitive myocardial infarction (MI) detector was proposed. A CNN model trained on 12-lead ECG signals was used for this purpose. In \cite{jun2018ecg}, a 2D CNN was proposed for predicting arrhythmia using transformed ECG signals to 2D grayscale images. In \cite{zhai2018automated} a subject-specific CNN was proposed for arrhythmia classification from ECG signals. The signals were transformed into a 2D matrix which preserved their temporal and morphological features. The model was trained on the most representative heartbeats for improving the performance. In \cite{ullah2021hybrid}, two ECG classifiers were presented using 1D and 2D CNN models. In this work, the 2D CNN model trained on ECG signals transformed to 2D images showed better results than the 1D CNN.

In \cite{ubeyli2009combining}, an ECG classifier was proposed in three phases, including an eigenvector feature extractor, a feature selector, and an RNN, which was trained on the extracted features. In \cite{ubeyli2010recurrent}, a four-class ECG classifier was presented using RNNs. The ECG signal's features were first extracted, and then the classification was made using selected features as inputs to the RNN model. In \cite{zhang2017patient}, a patient-specific RNN based ECG signal classifier was presented. Some morphology features of ECG signals were fed into RNN for better performance. In \cite{salloum2017ecg}, various RNN-based architectures were evaluated for ECG classification, and authentication with no need for ECG extracted features. In \cite{lynn2019deep}, a bidirectional gated RNN (BGRU) was introduced for ECG signal identification. The benefit of their proposed method was the network's ability to see both the past and future signal time steps simultaneously, which allowed for a better understanding of the ECG representations. Lastly in \cite{wang2019global}, an ECG feature extractor was proposed named Global RNN (GRNN). Every input ECG signal feature was extracted, and the most informative samples were selected by the optimization mechanism and set as the model's training data for better generalizability.

\subsection{Self-supervised ECG Representation Learning}

With the recent progress in self-supervised learning in areas such as computer vision \cite{chen2020simple} and natural language processing \cite{lan2019albert}, a number of recent works have utilized such methods for ECG representation learning. In what follows, we briefly present some prior work in this area. In \cite{sarkar2020self, sarkar2020selff}, self-supervised ECG-based emotion recognition was proposed to improve the network's performance in comparison to fully supervised learning. In this work, spatiotemporal ECG representations were learned by applying various transformations to unlabelled ECG signals, followed by predicting the applied transformation. Next, the downstream task was performed by freezing the pre-trained encoder component of the model and training a few dense layers with labelled data. In \cite{cheng2020subject}, a contrastive self-supervised learning approach was proposed for bio-signals. They tackled the inter-subject adverse effects on learning by presenting subject-aware optimization using a subject distinctive contrastive loss and adversarial training. 

In \cite{kiyasseh2021clocs}, a framework called CLOCS was proposed, in which a patient-specific contrastive learning method was able to achieve state-of-the-art results by learning the spatiotemporal representations of ECG signals. The method also benefited from the physiological features of ECG signals, namely temporal and spatial invariance to produce additional positive pairs from each patient signal. A model called TS-TCC was proposed in \cite{eldele2021time} which learned time-series data representations by creating two views of each sample by applying a strong and weak data augmentation, and then learning robust temporal features by applying a cross-view prediction task. In another method called 3KG proposed in \cite{gopal20213kg}, ECG representations were learned in a contrastive manner using physiological characteristics of ECG signals. 12-lead signals were transformed to a 3D space followed by application of augmentations in that space. The method was evaluated by finetuning on different heart diseases and produced strong results. Lastly, an approach called CLECG which was proposed in \cite{chen2021clecg} presented a contrastive self-supervised learning framework for ECG signals. The method utilized random cropping and wavelet transformations for contrastive learning augmentations.

In the end, we conclude from our review that self-supervised methods for ECG representation learning provide very promising results in comparison to fully supervised approaches. Nonetheless, the difficulty in labeling large amounts of data by experts poses open challenges in the area, and thus self-supervised solutions including contrastive learning are projected to further dominate the field. However, while the choice of encoder and classifier architectures in self-supervised frameworks is generally not too critical, the choice of the type and intensity of pretext or contrastive augmentations plays a critical role in ECG representation learning. This motivates our work presented in this paper where we provide a detailed analysis of different augmentations for ECG representation learning.

\section{Method}
In this section, we first explain our self-supervised contrastive framework shown in Figure \ref{fig:framework}. We then describe each of the augmentations that are applied to the ECG signals. Lastly we present the architectural details of different components of the pipeline.

\subsection{Self-Supervised Contrastive Framework for ECG Representation Learning}
Assume our input ECG signal as $x \in \mathbb{R}^{D \times L}$, where $D$ is the number of ECG leads, and $L$ is the length of each signal. In the first step, we apply an augmentation on $x$ and obtain the augmented sample $\tilde x$. The details of these augmentations will be explained in the next sub-section. Accordingly, we can assume $\tilde x$ as another view of $x$, and therefore, $x$ and $\tilde x$ are called positive pairs. Also, $x$ and $\tilde x$ make negative pairs with other input data and their respective augmentations. For better understanding, consider two different input ECG signals as $x_a$ and $x_b$, and their augmentations $\tilde x_a$ and $\tilde x_b$. In this example, the positive pairs are ($x_a$, $\tilde x_a$) and ($x_b$, $\tilde x_b$), while the negative pairs are ($x_a$, $x_b$), ($x_a$, $\tilde x_b$), and ($x_b$, $\tilde x_a$). 

\begin{figure}
 \begin{center}
 \includegraphics[width=1\columnwidth]{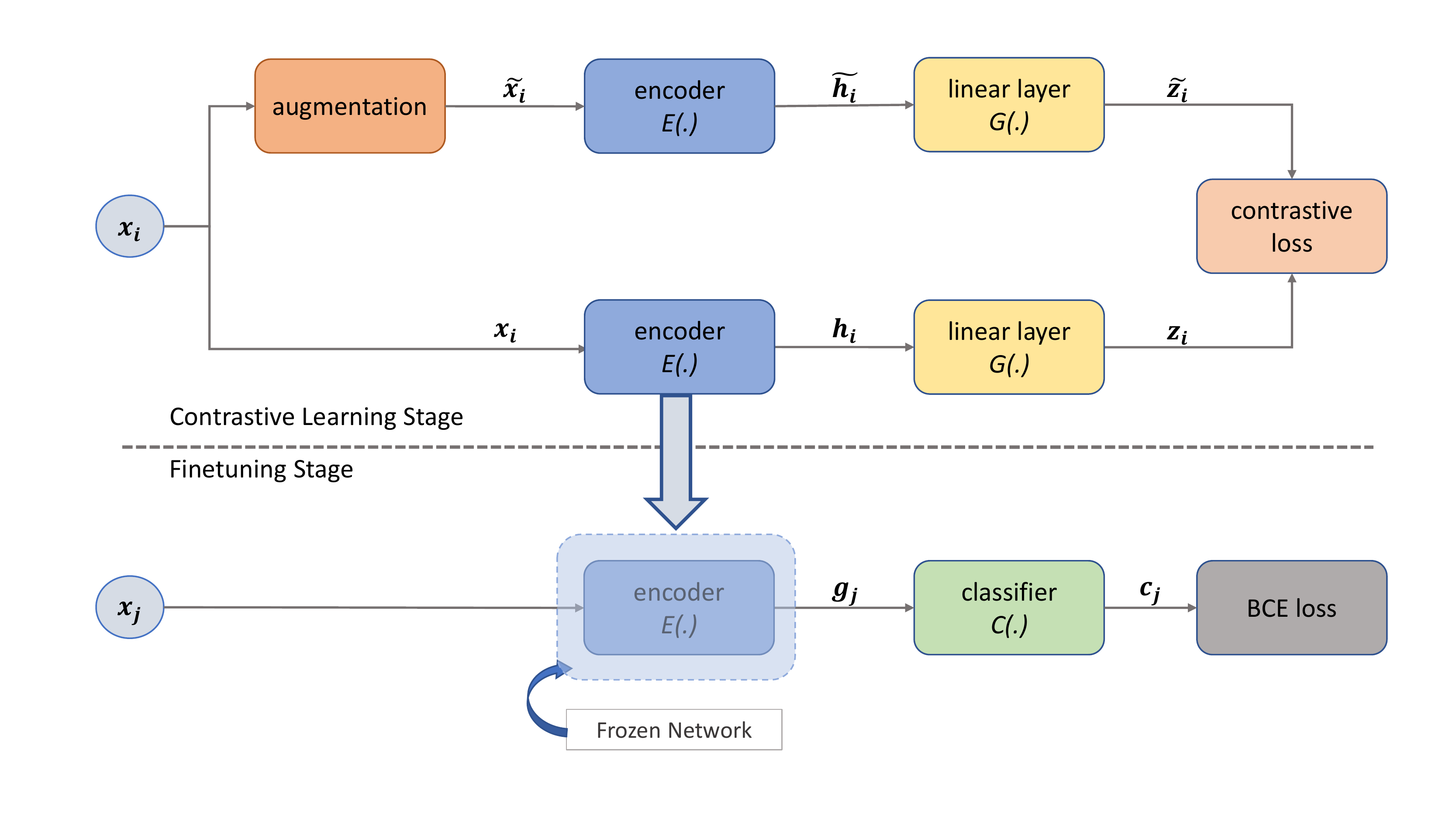}
 \end{center}
\caption{The contrastive framework used in our study.}
\label{fig:framework}
\end{figure}

Next, input signal $x_i$ and its augmented version $\tilde x_i$ are separately passed to our encoder model referred to as $E(.)$ through the two separate branches of our pipeline (see Figure \ref{fig:framework}), and result in $h_i$ and $\tilde h_i$. Next, $h_i$ and $\tilde h_i$ are passed through linear fully connected layers $G(.)$ for better characterization. The final outputs are $z_i$ and $\tilde z_i$, which are passed to the loss function for calculating the overall model contrastive loss. We use the loss presented in \cite{oord2018representation, gutmann2010noise} for the contrastive training stage of our framework. This loss function maximizes the agreement between different views of each sample, using the variations among different samples. The loss function is formulated as:
\begin{equation}
    L_{contr} = -log(\frac{exp(sim(z_i,\tilde z_i)/\tau)}{\sum_{k=1} ^{2N} \mathbbm{1}_{(i\ne k)}exp(sim(z_i,z_k)/\tau)}),
\end{equation}
where $N$ is the batch size and $\tau$ is the temperature parameter. The $sim(z_i,\tilde z_i)$ function used in the loss formula is the cosine similarity with the formulation of:
\begin{equation}
    sim(z_i,\tilde z_i)=\frac{z_i.\tilde z_i}{\parallel z_i\parallel \parallel \tilde z_i\parallel} .
\end{equation}
Following some previous works \cite{chen2020simple, gopal20213kg}, in a batch of size $N$, only each sample and its augmented form are selected as positive pairs, while the other $2N-2$ samples are selected from negative pairs.

After contrastive training of the network, we freeze the weights of $E(.)$ and transfer them to another pipeline for the downstream task. In this step, $G(.)$ is no longer required and thus discarded. Please refer to Figure \ref{fig:framework}. 
In this phase, we use labelled samples $(x_j, l_j)$ where $x_j$ is the sample and $l_j$ is its label. $(x_j, l_j)$ is fed to the pretrained $E(.)$ followed by a few untrained fully connected layers $C(.)$. The goal of these layers is to perform the final classification and $c_j$ are the predicted classes. The network is then trained by supervised learning with Binary Cross-Entropy (BCE) loss. The formula for the BCE loss is as follows:
\begin{equation}
     L_{BCE} = -\frac{1}{N} \sum_{i=1} ^{N} [l_i\times log(c_j)+(1-l_j)\times log(1-c_j)],
\end{equation}
where $N$ is the total number of fine-tuning samples in each batch, $l_j$ is sample's true label and $c_j$ is the model output for the $j^{th}$ class.

\begin{figure*}
    \centering
    \includegraphics[width=.96\linewidth]{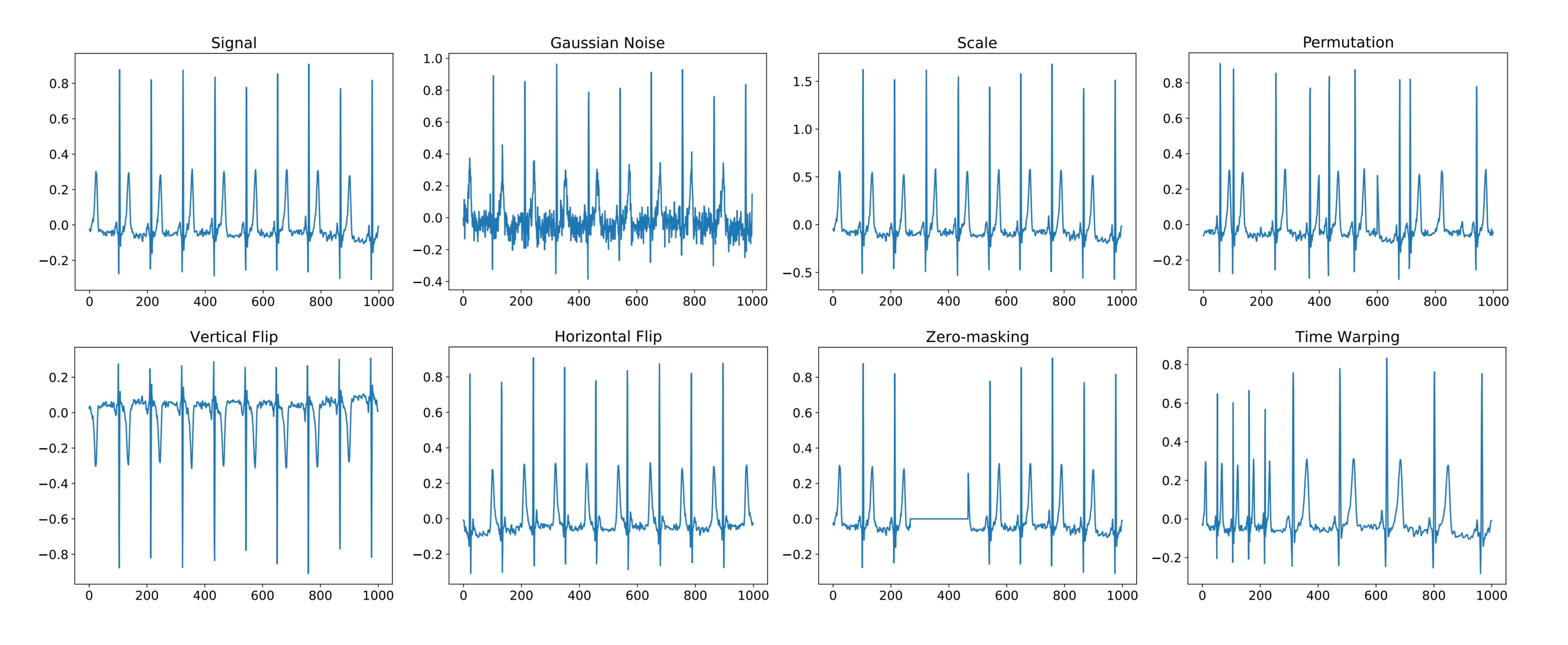}
    \caption{Different augmentations applied on a sample ECG segment.}
    \label{fig:augmentations}
\end{figure*}

\subsection{Augmentations}

As observed and discussed earlier in Section 2, in self-supervised and contrastive approaches, the type and intensity of different augmentations used to learn ECG representations in the contrastive learning step can play a critical role in downstream classification. Here, to conduct a detailed and rigorous study on this aspect of our contrastive learning framework, we carefully select a large number of augmentations that have been shown in different literature to help with generalization in ECG learning. Some of these augmentations very closely resemble artifacts that can appear in the wild or even in the lab when acquiring ECG signals, and therefore their use in contrastive self-supervised learning can help the model extract effective and used representations and generalize to unseen data. For instance the addition of Gaussian noise can occur in various scenarios when recording ECG data. As another example, ECG can be scaled due to changes in sensor conductance that can occur naturally. As another example, ECG samples may also be dropped during collection. In addition to the impact of the type of augmentation, their intensities can also greatly impact the outcome. Prior works have shown that generally during self-supervised learning, when augmentations are very weak, poor representations are learned since the augmented signals are very similar to the input samples. On the other hand, it has been also shown that when augmentations are too strong, the augmented samples are so different from the original signals that the network does not learn useful and generalizable representations. As a result, we use eight different augmentation techniques in our framework and systematically cover a wide range for their respective parameters. The augmentation details are as follows.

\subsubsection{\textbf{Noise}}
We add Gaussian noise $N(t)$ to the ECG signal $x(t)$. This is formulated as:
$P(x)=\frac{1}{\sqrt{2\pi\sigma}}e^{-\frac{(x-\mu)^2}{2\sigma^2}}$, where $\mu$ is the mean value set to zero, and $\sigma$ is the standard deviation for our augmentations.
The augmented ECG signal is computed by adding the noise signal to the ECG as $\tilde{x}(t) = x(t)+N(t)$. 

\subsubsection{\textbf{Scale}}
The input ECG is multiplied by a scaling factor $S$ to obtain $\tilde{x}(t) = S\times x(t)$.

\subsubsection{\textbf{Permutation}}
In this augmentation, the ECG signal 
is divided into $m$ sub-sections, ${x_1, x_2, ..., x_m}$. These sub-sections are then shuffled and concatenated together to make the augmented signal $\tilde{x}$. 

\subsubsection{\textbf{Vertical Flip}}
The input signal is flipped vertically across the time axis as $\tilde{x}(t) = -x(t)$.

\subsubsection{\textbf{Horizontal Flip}}
To perform this augmentation, we modify $x(t)$ as $\tilde{x}(t) = x(-t+L)$ where $L>0$ is the length of $x(t)$.

\subsubsection{\textbf{Zero Masking}}
For this augmentation, a masking factor $r$ is selected which determines the percentage of the original signal $x(t)$ that will be set to zero. Next, we randomly set $r \times L$ consecutive samples of $x(t)$ to zero to obtain $\tilde{x}(t)$.

\subsubsection{\textbf{Time Warping}}
For this augmentation, we divide $x(t)$ into $m$ segments $x_1(t), \cdots, x_m(t)$ where $m$ is even. Next, we use time warping to stretch half of the segments (randomly selected) by $w\%$ while simultaneously squeezing the other half by the same factor. The segments are finally concatenated in the same order to obtain $\tilde{x}(t)$ where the final length of $\tilde{x}(t)$ remains unchanged as $L$.

Figure \ref{fig:augmentations} illustrates an input ECG signal, as well as samples of each of the augmentations mentioned above. We can observe that each augmentation alters the spatiotemporal properties of the signal in a unique way, overall resulting in a wider distribution of shapes and characteristics, which can lead to more generalized learning.

\begin{figure*}
    \centering
    \includegraphics[width=0.7\linewidth]{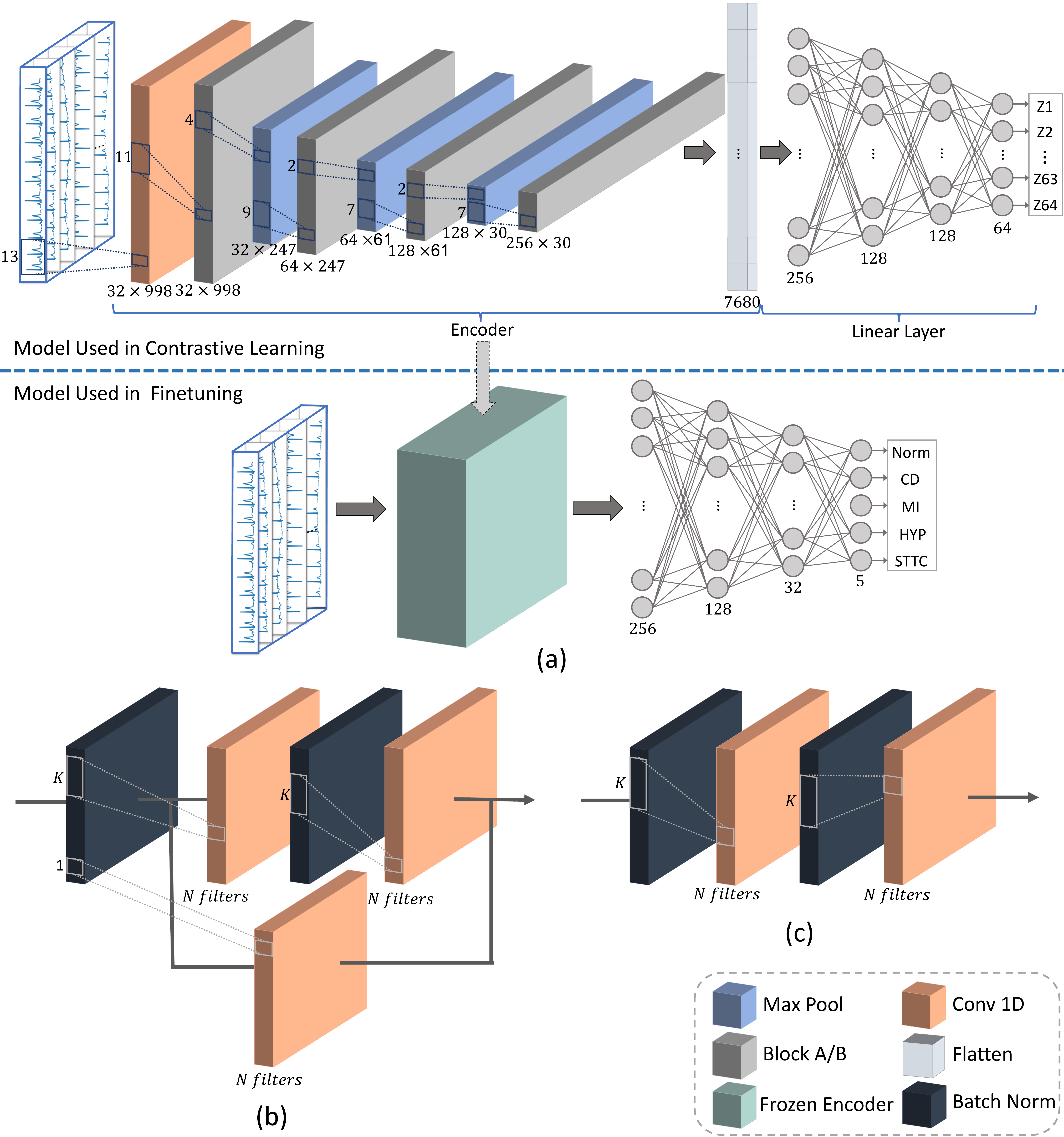}
    \caption{The architecture details of (a) encoder and linear layers for contrastive training and fine-tuning, (b) the block architecture of Encoder A, and (c) the block architecture of Encoder B.}
    \label{architecture}
\end{figure*}

\subsection{Model Details}
As discussed earlier in Section 3.A, our framework uses a deep encoder, $E$, for extracting ECG features from raw data. In order to explore the flexibility and generalizability of our proposed model, we use two different CNN architectures for $E$. The first model is based on residual network adopted from \cite{cheng2020subject}, which has shown great performance in bio-signal classification. Henceforth, we refer to this network as \textit{Encoder A}. As shown in Figure \ref{architecture}(a) and (b), this encoder contains one 1D convolutional layer, four residual blocks (ResBlocks), three max-pooling layers, and a Flatten layer. Each ResBlock consists of two 1D convolutional layers with ELU activation function, two batch normalizations, and a 1D convolutional layer with a residual connection. The numbers in the convolutional blocks of Figure \ref{architecture} represent the number of input channels, output channels, and kernel sizes respectively.

A flatten layer as the final step of the encoder, which outputs a final feature vector with the size of $1\times 7680$. The second encoder, \textit{Encoder B}, consists of layers similar to \textit{Encoder A}, except for the residual connections inside the convolutional blocks. The details can be seen in Figure \ref{architecture}(a) and (c).

Following the encoder, we use four feed-forward linear layers with ReLU activation functions. 
The numbers in the linear layers in Figure \ref{architecture} (a) represent the input and output sizes respectively.
 
In the end, as described earlier in Section 3.A, after training the contrastive model, the encoder is frozen and a classification model is added and fine-tuned. This classifier consists of four fully connected layers with ReLU activation functions. The details of this stage of the model is presented in Figure \ref{architecture}(a).

\begin{table}
    \caption{The details of the augmentation parameters used for contrastive learning.}
    \label{table_1}\centering
    \begin{tabular}{|c|c|c|c|}\hline
    \textbf{Augmentation Method} & \textbf{Min} & \textbf{Max} & \textbf{Data Points} \\\hline\hline
    Gaussian Noise ($\sigma$)  & 0.01 & 1 &[0.01, 0.03, 0.05, 0.07, 0.1,\\&&&0.15, 0.2, 0.25,0.4, 0.6, 0.9]\\\hline
    Scale ($S$) & 0.1 & 3 & [0.1, 0.3, 0.5, 0.8, 1.2,\\&&&1.7, 2, 2.5, 3]\\\hline
    Permutation & 2 & 20 & [2, 4, 5, 8, 10, 20]\\\hline
    Horizontal Flip & --  & --  & --\\\hline
    Vertical Flip & --  & --  & --  \\\hline
    Zero Mask ($r$) & 10\% & 60\% & [10\%, 20\%, 30\%, 40\%,\\ & & &50\%, 60\%]\\\hline
    Time warping & 0.25 & 0.75 & [2-(0.25, 0.5, 0.75),\\&&&4-(0.25, 0.5, 0.75)]\\\hline
    \end{tabular}
\end{table}

\begin{figure}
 \begin{center}
 \includegraphics[width=.93\columnwidth]{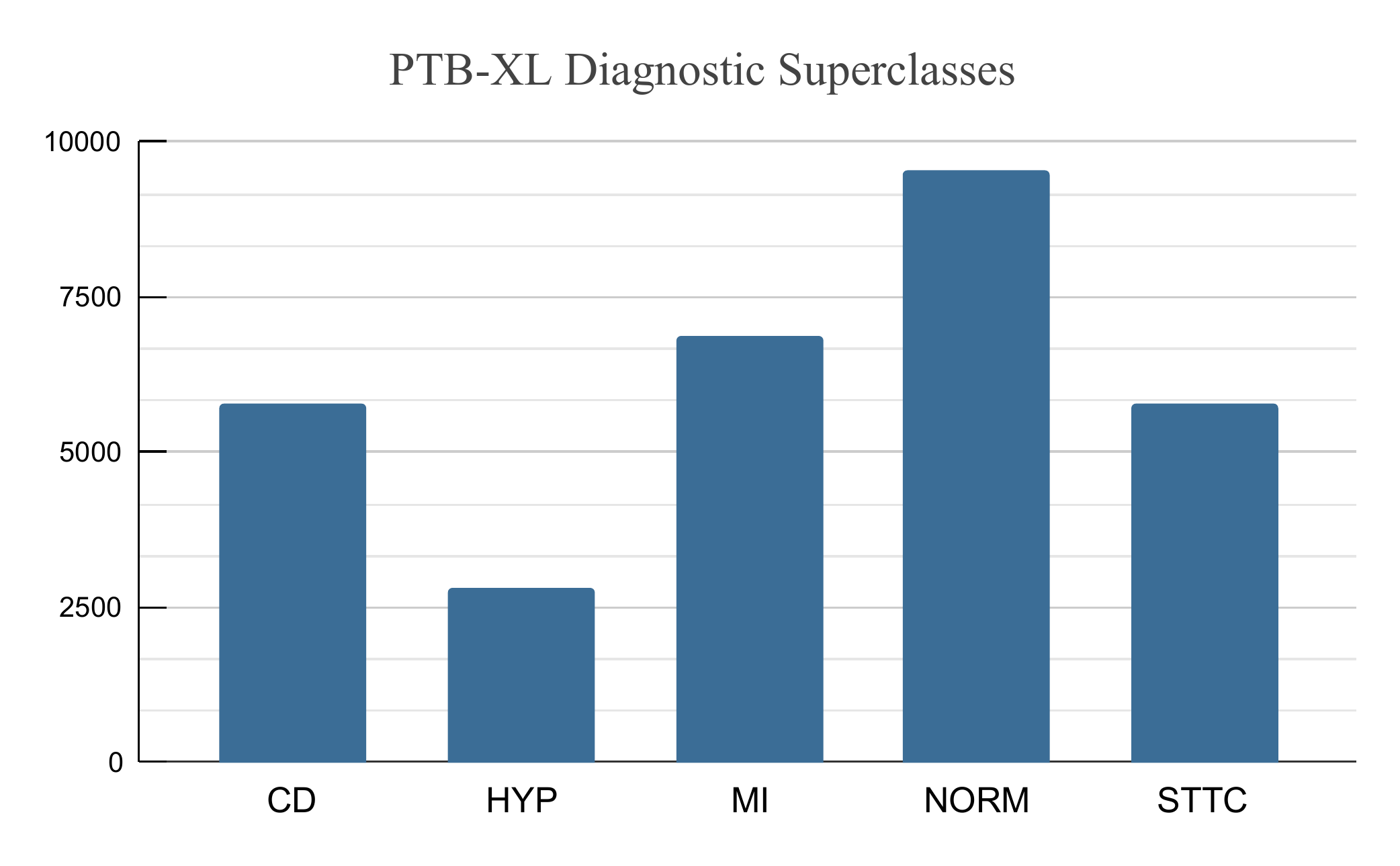}
 \end{center}
\caption{The distribution of classes in the PTB-XL dataset.}
\label{fig:ptbxl}
\end{figure}

\begin{figure*}[!ht]
    \centering
    \includegraphics[width=0.8\linewidth]{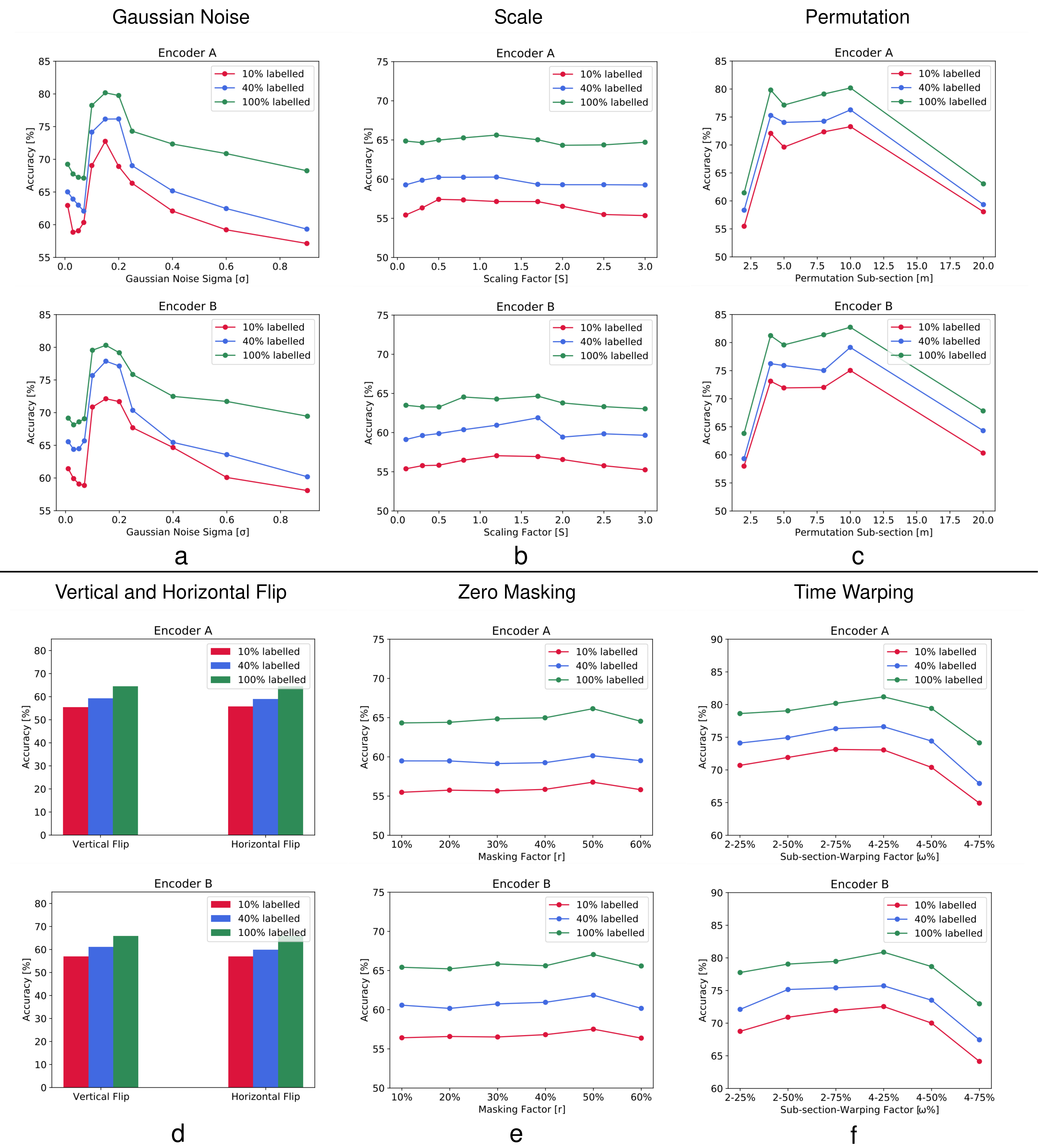}
    \caption{Accuracy of our contrastive self-supervised model for different augmentation parameters: (a) adding Gaussian noise, (b) scaling, (c) permutating, (d) vertical and horizontal flipping, (e) zero masking, and (f) time warping. Two encoders, A (with residual connections) and B (without residual connections), have been used in these experiments.}
    \label{fig:accuracy}
\end{figure*}

\section{Experiments and Results}
\subsection{Dataset}
We use a large publicly available dataset, PTB-XL \cite{wagner2020ptb}. This dataset consists of 21837 samples of 12-lead ECG from 18885 patients collected with Schiller AG devices. The dataset is categorized into five main classes, normal, myocardial infarction, conduction disturbance, ST/T changes, and hypertrophy. Figure \ref{fig:ptbxl} shows the distribution of these five classes in detail. The dataset is provided with two sampling frequencies, 100 and 500 \textit{Hz}. In our work, we used signals with a frequency of $f_{s}$=100 \textit{Hz} to reduce our computational requirements. The dimension of the ECG samples fed into our model is $12\times 1000$, corresponding to 1000 points that is 10 seconds long and 12-leads.

\subsection{Training Details and Evaluation}
PTB-XL dataset is initially separated into ten subsections, and by using these subsection, we split the dataset into two main subsections, 10\% for testing the model's performance, and 90\% for contrastive training and finetuning.  
For evaluating the effects of the number of labelled data in finetuning the model, each model is finetuned three times using different amounts of training data, $\frac {1}{9}$, $\frac {4}{9}$ and $\frac {9}{9}$, henceforth approximated and referred to as 10\%, 40\%, and 100\%.

We use Adam optimizer~\cite{kingma2014adam} with a learning rate of $5\times10^{-4}$ and weight decay of $1\times10^{-3}$ to train the contrastive framework. The contrastive stage is trained for 50 epochs with a batch size of 128. In the downstream phase, the added linear layers are trained for 200 epochs using Stochastic Gradient Descent (SGD) optimizer with a learning rate of 0.01. We use weighted accuracy to evaluate the performance of our model. The method is implemented using PyTorch and trained on NVIDIA GeForce GTX 1070 GPU.

\subsection{Results}
This section illustrates our experimental results on the PTB-XL dataset. As comprehensively described earlier in section 3.A, we apply seven different augmentations on ECG signals. For each augmentation, we use various parameters as shown in Table \ref{table_1}. To examine the efficiency of Gaussian noise adding to ECG signal, we use Gaussian noise with zero mean and $\sigma$ in the range of $0.01-0.9$, with a step size of $0.05$. We use scale augmentation with a scaling factor of $0.1-3$ and a step size of $0.2$. The zero-masking augmentation is applied on the ECG signals with a range of $10\%-60\%$ and a step size of $10\%$. For permutation augmentation, a range of $2-20$ is selected with a step size of $2$. Two different variations are used for time warping, one where the signal is segmented into 2 pieces, while in the other, the signal is divided into 4 segments. We apply 3 different scaling factors for each variation, $25\%$, $50\%$, and $75\%$. We also flip the ECG signals both horizontally and vertically. The results for each of these augmentations are shown in Figure~\ref{fig:accuracy}. We observe that as expected, different augmentations and different parameters result in varied accuracies. We further train the inference model (Figure~\ref{architecture} (a) finetuning) in a fully supervised manner with all the training labels available. As we dive into each augmentation result in detail in the following paragraphs, the fully supervised results shown in Figure~\ref{fig:accuracy} provide a valuable benchmark in understanding the impact of each augmentation. We should point out that while the performance of self-supervised learning is expectedly lower than the fully supervised counterpart in most cases, it is significantly less reliant on the labels during training as only the classifier portion is trained using the original labels.

In Figure~\ref{fig:accuracy} (top left) we observe that augmenting via Gaussian noise results in learning strong ECG representations in the range of $\sigma=$ $0.1$ to $0.25$, while outside of this range, performance drops. 
This is due to the fact that when $\sigma=[0.01-0.07]$, the augmentation is likely too weak, resulting in augmented ECG signals that are very similar to the original signals. The network therefore does not learn effectively. On the other hand, for $\sigma=[0.4-0.9]$, we hypothesize that the augmentation is too strong, and thus the augmented ECG signal looks very different from the original signal. Therefore, associating the two signals as the same sample becomes too difficult for the network, and thus learning does not take place effectively. Next, we observe that a similar trend is observed when different amounts of labels ($10\%$, $40\%$ and $100\%$) are used during finetuning. However, while the use of larger amounts of labelled data results in better overall performance, our model does not suffer considerably when the amount of labels are reduced by approximately $90\%$. Lastly, both Encoder A and Encoder B exhibit similar performances, which point to the generalizability of our approach.

Figure~\ref{fig:accuracy} (top middle) demonstrates that the scaling augmentation on ECG signals does not show overall promising results in comparison to the other applied augmentations. Nevertheless, the results slightly improve when scaling with the scaling factor of $S=[0.5-1.7]$. We believe that scaling does not show favourable results given that this augmentation is too simple, and does not apply sufficient spatiotemporal changes to the signal for the augmented samples to expand the learned distribution for better representation learning. We further observe that varying amounts of labelled data used for finetunig increases the performance by approaximately $10\%$, yet, the results are still not very strong and only a very weak trend can be seen with respect to the scaling factor. Moreover, the choice of the encoder does not considerable affect the performance, pointing to the fact that this augmentation, at least when used individually, is likely not very helpful towards self-supervised contrastive ECG learning.

We observe from Figure~\ref{fig:accuracy} (top right) that when the permutation augmentation is applied on ECG signals, using $2$ and $20$ sub-sections does not show promising results. We believe that using $2$ sub-sections will result in augmentations that are too simple to learn for the model, while using $20$ sub-sections becomes extensively challenging to learn. As mentioned earlier, each ECG sample is $10$ seconds long in our experiments. On average, the resting human heart rate is between $60$ and $100$ beats per minute \cite{quer2020inter}, which means that our ECG samples are likely to have $6$ to $10$ heartbeats per sample. Therefore, when we divide these samples into 20 sub-sections and shuffle them randomly, it is likely that the resulting augmented sample does not resemble an ECG signal anymore, which is why the network is not able to learn ECG representations, and thus poor performance. Similar to the Noise augmentation which also proved to be effective, we observe that reduction of labelled data for finetuning, even by the significant amount of $90\%$, does not hurt the performance considerably. Moreover, the choice of encoder does not hugely impact the results. In the end, we conclude that by and large, this augmentation is effective for contrastive ECG learning.

The results shown in Figure \ref{fig:augmentations} (bottom left) indicate that vertical and horizontal flip, on the other hand, are not effective augmentations. We hypothesize that the spatiotemporal properties of the resulting signals are different enough from the original samples that the network is unable to learn. The behaviour of the model when trained with these augmentations is similar to other augmentations in terms of the use of labelled samples or encoder. As seen in Figure \ref{fig:augmentations} (bottom middle), also does not show promising results regardless of the choice of encoder or the amount of labelled data. We do observe a small peak in the accuracy with $L=50\%$, but overall, this augmentation does not seem to allow the network to learn effective ECG representations. Further details on the limitations of our experiment design in this regard and how to potentially improve the model to benefit from this augmentation is discussed below in the \textit{Limitations} section.

Finally, Figure \ref{fig:augmentations} (bottom right) presents the results of the time warping augmentation, which shows promising performance. We observe that when we divide the ECG signal into 2 segments and then apply time warping, the stronger squeeze and stretch factor achieves better results. However, when dividing the signal into 4 segments, the weaker squeeze and stretch factor shows better performance. We believe that for fewer ECG segments, i.e. 2, for the model to learn better representations, we need a stronger squeeze and stretch factor so that the augmented signal does not become too easy for the model to learn. Yet for more segments, the smaller squeeze and stretch factor works better since the augmentation is already not so simple and it is not reasonable to add even more complexity by using the stronger squeeze and stretch factor.

In the end, in order to obtain a high-level view of the augmentations in comparison to one another, Table \ref{table:results_compare} illustrates the best results obtained for each augmentation and for different amounts of labelled data used in finetuning. We also train both our models in a fully supervised manner to achieve a better understanding of the effectiveness of contrastive training. For the model with Encoder A we achieve an accuracy of $82.428$, and for the model with Encoder B, we get $84.876$ accuracy. From this analysis, we observe that adding Gaussian noise, permutation, and time warping show the best performances for contrastive ECG learning, while scaling, zero masking, and vertical/horizontal flipping show lower performances.

\begin{table}[htbp]
\caption{Comparison of the best accuracies achieved with Encoder A (residual) and Encoder B (non-residual) using different augmentations and finetuned with 10\%, 40\%, and 100\% labelled data. }
\centering
\begin{tabular}{|l|l c c|} \hline 
\textbf{Augmentation} & \textbf{Labelled Data} & \textbf{Model A} & \textbf{Model B}\\ \hline\hline
     & 10\% & 68.91 & 72.13\\
    Gaussian Noise & 40\% & 76.17 & 77.87\\
     & 100\% & 80.17 & 80.31\\
    \hline
  & 10\% & 57.41 & 57.04\\
    Scale & 40\% & 60.27 & 61.89\\
     & 100\% & 65.62 & 64.65\\
    \hline
    & 10\% & 73.27 & 75.04\\
    Permutation & 40\% & 76.27 & 79.14\\
     & 100\% & 80.19 & 84.73\\
    \hline
    & 10\% & 55.47 & 57.03\\
    Vertical Flip & 40\% & 59.28 & 60.16\\
     & 100\% & 64.45 & 65.83\\
    \hline
    & 10\% & 55.66 & 56.98\\
    Horizontal Flip & 40\% & 58.94 & 58.03\\
     & 100\% & 64.50 & 65.94\\
    \hline
    & 10\% & 56.77 & 57.51\\
    Zero Masking & 40\% & 60.14 & 61.86\\
     & 100\% & 66.14 & 67.05\\
    \hline
    & 10\% & 73.13 & 72.54\\
    Time Warping & 40\% & 76.61 & 75.72\\
     & 100\% & 81.18 & 80.87\\
    \hline
\end{tabular}
\label{table:results_compare}
\end{table}

\subsection{Limitations}
Our study has a number of limitations which will be explored in future work. \textbf{(\textit{i})} One approach in expanding out work will be to explore smaller step sizes and larger ranges in our search for the optimum parameters for the augmentations. For instance, the zero-masking augmentation can be modified to allow for shorter zero-masked segments distributed throughout the ECG signal. \textbf{(\textit{ii})} Moreover, more advanced augmentations, for example those in the frequency domain, could be considered. \textbf{(\textit{iii})} To further explore the generalizability and practicality of our study, additional datasets as well as larger number of output classes will be trained and evaluated against. In particular, a study of augmentations that aid in classification of various types of arrhythmia will be conducted to allow for arrhythmia-specific findings. \textbf{(\textit{iv})} In this study, we explored each augmentation individually. Nonetheless, combinations of augmentations may exhibit different behaviors, which we will explore through forward/backward selection methodologies. \textbf{(\textit{v})} Lastly, additional variations of contrastive learning can be studied and explored for identifying the optimum set of augmentations.

\section{Conclusion}
This paper highlights the significance of augmentation selection for ECG representation learning using contrastive self-supervised learning. Our experiments are done on the PTB-XL, a large and public arrhythmia-specified dataset. In our experiments, the primary model is first trained with unlabelled augmented data contrastively to learn ECG representations, followed by freezing the encoder and finetuning a few linear layers with different amounts of labeled data to conceive the final results. Our experiments demonstrate that particular augmentation techniques result in better and more generalizable learning of ECG representations with some augmentations such as vertical/horizontal flipping showing poor performance. Moreover, we find that the range of augmentation complexities plays an important role in the performance as augmentations that are too weak or too strong do not result in effective training. Our study uncovers optimum ranges of complexities for different augmentations such as noise addition, time warping, and permutation, which can be used by researchers in the area for effective self-supervised ECG representation learning.

\printbibliography
\small
\end{document}